\documentclass{article}
\usepackage{ijcai19}

\usepackage{times}
\usepackage{soul}
\usepackage{url}
\usepackage[hidelinks]{hyperref}
\usepackage[utf8]{inputenc}
\usepackage[small]{caption}
\usepackage{graphicx}
\usepackage{amsmath}
\usepackage{booktabs}
\usepackage{algorithm}
\usepackage{algorithmic}
\usepackage{bm}
\usepackage{amsfonts}

\usepackage{epsfig}
\usepackage{tabularx}
\usepackage{enumerate}
\usepackage{subfigure}
\usepackage{makecell}
\usepackage{multirow}
\usepackage{epstopdf}
\usepackage{helvet}
\usepackage{courier}
\usepackage{hyperref}
\usepackage{color}
\usepackage{dsfont}

\title{K-margin-based Residual-Convolution-Recurrent Neural Network for Atrial Fibrillation Detection}

\author{
Yuxi Zhou$^{1,2}$\and
Shenda Hong$^{1,2}$\and
Junyuan Shang$^{1,2}$\and
Meng Wu$^{1,2}$\and
Qingyun Wang$^{1,2}$\and \\
Hongyan Li$^{1,2, }$\footnote{Contact Author}\And
Junqing Xie$^{3}$\\
\affiliations
$^1$School of Electronics Engineering and Computer Science, Peking University, Beijing, China\\
$^2$Key Laboratory of Machine Perception (Peking University), Ministry of Education, Beijing, China\\
$^3$Medical Informatics Center, Peking University, Beijing, China\\
\emails
\{joy\_yuxi, hongshenda, sjy1203, wumeng93, wangqingyun, leehy, xjq\}@pku.edu.cn
}

\begin{document}
\maketitle

\begin{abstract}
Atrial Fibrillation (AF) is an abnormal heart rhythm which can trigger cardiac arrest and sudden death. Nevertheless, its interpretation is mostly done by medical experts due to high error rates of computerized interpretation. One study found that only about $66\%$ of AF were correctly recognized from noisy ECGs. This is in part due to insufficient training data, class skewness, as well as semantical ambiguities caused by noisy segments in an ECG record. In this paper, we propose a K-margin-based Residual-Convolution-Recurrent neural network (K-margin-based RCR-net) for AF detection from noisy ECGs. In detail, a skewness-driven dynamic augmentation method is employed to handle the problems of data inadequacy and class imbalance. A novel RCR-net is proposed to automatically extract both long-term rhythm-level and local heartbeat-level characters. Finally, we present a K-margin-based diagnosis model to automatically focus on the most important parts of an ECG record and handle noise by naturally exploiting expected consistency among the segments associated for each record. The experimental results demonstrate that the proposed method with 0.8125 $F_{1NAOP}$ score outperforms all state-of-the-art deep learning methods for AF detection task by $6.8\%$. 
\end{abstract}

\section{Introduction}
Atrial Fibrillation (AF) is an abnormal heart rhythm characterized by rapid and irregular beating of the atria, which can trigger cardiac arrest and sudden death as a consequence of pumping blood less effectively \cite{ye2010arrhythmia,Janse2006Basis}. The Electrocardiograph (ECG) has been a cornerstone for the detection and diagnosis of such condition for a long time. However, its interpretation is mostly done by medical experts due to the high error rates of computerized interpretation. It has been found that good performance was shown on carefully-selected often clean data \cite{clifford2017af}, and only about 66\% of AF predictions were correctly recognized from noisy ECG data \cite{shah2007errors}. This is especially problematic in developing countries, where the availability of clinics and medical experts is low \cite{gradl2012real} \cite{silva2011improving}. Accordingly, there is a definite need for automatic, accurate physiological monitoring solutions for AF detection (without domain knowledge and features given by domain experts in advance), which can be used in home or ambulatory settings. 

\begin{figure}[tbp]
 \centering
 \includegraphics[width=8cm]{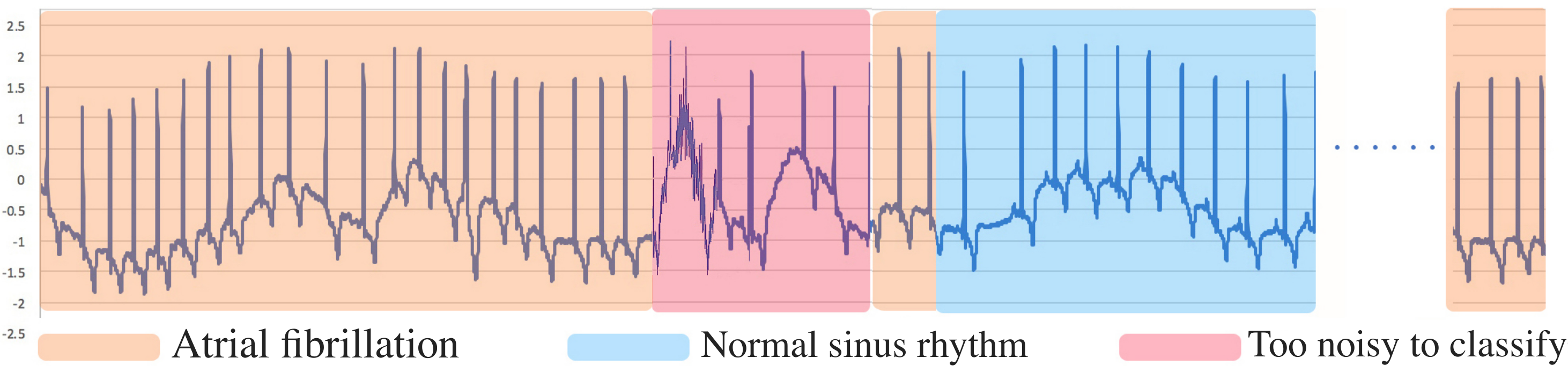}
 \caption{An example of an AF record with noisy segments.}
 \label{fig:af_intro}
\end{figure}

Despite the significance of this problem, it is challenging to reliably detect AF from noisy ECGs due to the following aspects. First, there is insufficient training data for AF detection, since well-labelled ECG recordings are hard to acquire. Second, the inevitable existence of class skewness in ECGs (as diseases happen rarely) may result in poor performance. Then, it is hard to capture both local heartbeat-level characters and long-term rhythm-level trend in an ECG record. Moreover, there may be semantical ambiguities caused by noisy segments in an ECG record as shown in Figure \ref{fig:af_intro}. Learning features from these noisy segments will lead to poor Deep Neural Network (DNN) models.

\begin{figure*}[tbp]
 \centering
 \includegraphics[width=0.96\textwidth]{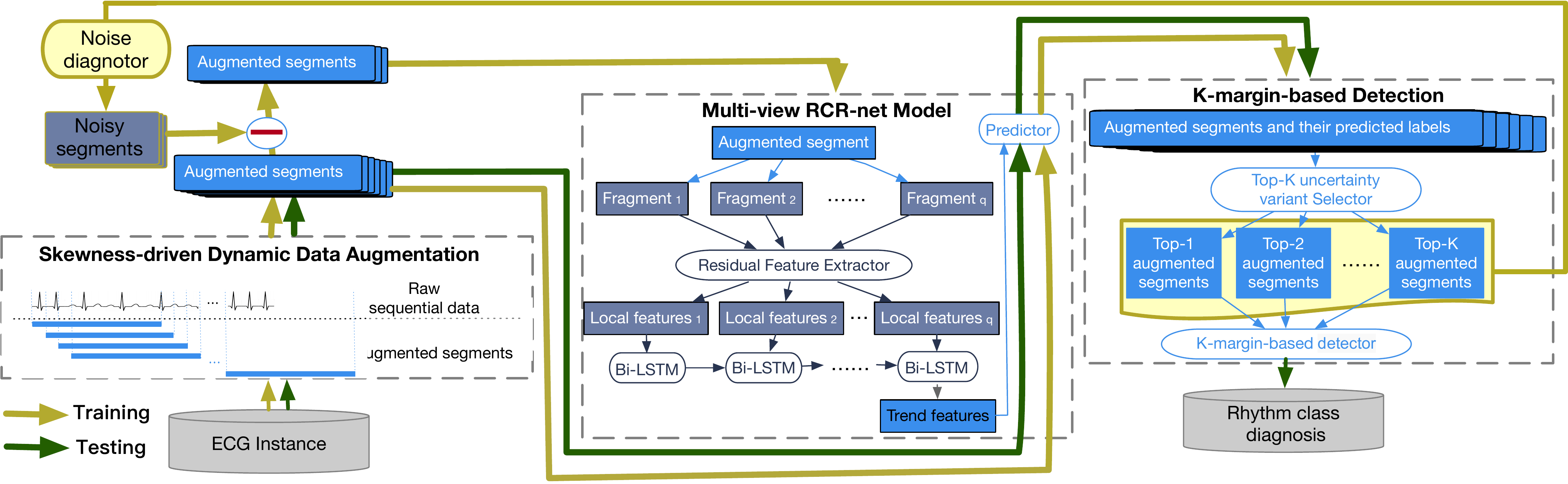}
 \caption{The framework of K-margin-based RCR-net model}
 \label{fig:framework}
\end{figure*}

In this paper, we propose a K-margin-based Residual-Convolution-Recurrent neural network (K-margin-based RCR-net) for AF detection from noisy ECGs. We first employe a skewness-driven dynamic augmentation method to handle problems of data inadequacy, class imbalance and high computing cost in ECG analysis. Then, we propose a multi-view learning method based on a novel Residual-Convolution-Recurrent neural networks (RCR-net) to automatically extract both long-term rhythm-level and local heartbeat-level characters. Besides, a K-margin-based diagnosis model is presented to identify the classification of each ECG by selecting its most probable label of relevant top-K augmented segments with the least certainty margin, which can automatically focus on the most important segments associated with an ECG record and handle noise by naturally exploiting expected consistency among the segments associated for each record.
We carry out thorough experiments on PhysioNet/Computing in Cardiology Challenge 2017 Dataset \cite{clifford2017af}. The experimental results demonstrate that the proposed method with 0.8125 $F_{1NAOP}$ score outperforms all state-of-the-art deep learning methods for AF detection task by $6.8\%$

\section{Related Work}
AF detectors can be divided into two categories including feature engineered methods and deep learning based methods. 

The feature engineered methods are highly related to cardiological knowledge which can achieve high accuracy if the recorded ECG signals are clean and high-resolution. In AF detection task, a process of using cardiological knowledge to extract informative values is essential to make classifiers able to distinguish AF patients from the population. Although engineering effective features is difficult and time-consuming, there are many excellent works that proposed various kinds of features and got good detection performance \cite{tateno2001automatic,linker2016accurate,carrara2015heart}. These methods help cardiologist to promote medical service quality in hospital. However, the biggest problem is that it's hard to extract accurate features from noisy contaminated ECGs. 

Recently, deep learning-based methods like Convolutional Neural Network (CNN) \cite{kiranyaz2015real,xiong2017robust,andreotti2017comparing,chandra2017atrial,kamaleswaran2018robust,hannun2019cardiologist}, Recurrent Neural Network (RNN) \cite{schwab2017beat}, Convolutional Recurrent Neural Network (CRNN) \cite{zihlmann2017convolutional}, and combined methods \cite{hong2017encase,hong2019} have achieved success in AF detection due to their powerful automatic feature learning ability. 
However, as far as we are concerned, these methods still have high error rates for ECG diagnosis when encountering insufficient training data, class skewness, as well as the semantical ambiguities of noisy ECGs.

\section{Methods}
\subsection{Problem Definition and General Framework}
AF detection is the task of automatically classifying an ECG into one of cardiac arrhythmia classes. Formally, we denote the training dataset as $\mathcal{D} = \{X,Y\}$ where $X=[\bm{x}^{(1)}, \bm{x}^{(2)}, \dots, \bm{x}^{(N)}]$ are ECG sequence inputs, $Y = [y^{(1)},y^{(2)},\dots,y^{(N)}]$ the corresponding label set and $N$ the total number of training data. Given the labelled training dataset $\mathcal{D}$, our goal is to learn a predictive model which takes an unlabelled ECG sequence $\bm{x}^{(i)}$ as input and outputs the prediction $\hat{y}^{(i)} \in \mathcal{C}$ where $\mathcal{C}=\{c_1, c_2, \dots, c_m\}$ is a set of $m$ different rhythm classes.





In this paper, we consider a DNN model as the basic sequence classifier. Recent deep learning works have demonstrated significant success on sequential classification tasks, that can naturally integrate and extract hierarchy features automatically. For the purpose of benefiting from both automatic feature extraction from CNN and capturing the long-term trend from RNN, we employ a Residual-Convolutional-Recurrent Neural Network (RCR-net) as shown in Figure \ref{fig:framework}. 

Nevertheless, a RCR-net model still can't be directly applied for reliably detecting AF from noisy ECG segments. On the one hand, ECGs are generated continuously with hundreds of millions of points of each patient, training a DNN model on such a long time period may result in high computing complexity. On the other hand, the class skewness and semantical ambiguities caused by noisy ECG segments may lead to poor, or even unacceptable quality of DNN models. 

To solve the aforementioned problems, we adapt a K-margin-based learning approach. The framework of K-margin-based RCR-net is shown in Figure \ref{fig:framework}. In detail, we firstly preprocess raw data using Skewness-driven Dynamic Data Augmentation (Sec.~\ref{sec:augment}) to relieve the class imbalance problem. Then, the RCR-net model (Sec.~\ref{sec:rcr-net}) is trained using the labelled augmented segments. It is noteworthy that only a portion of the segments in each record participate in the RCR-net learning process. Finally, a K-margin-based diagnosis model (Sec.~\ref{sec:build}) is employed to identify the classification of each ECG record by automatically focusing on the most important segments and handling noise. 

\subsection{Skewness-driven Dynamic Data Augmentation} \label{sec:augment}
As introduced before, to solve or relieve the problems of inadequate data, classes skewness and high computing complexity, we propose a skewness-driven dynamic augmentation method to handle these problems. Specifically, we deploy slide-and-cut methodology to generate more short-term ECG segments based on original long ECG records. Usually, slide-and-cut needs two predefined parameters: 1) \textit{windows\_size} (denoted as $w$): length of cut, 2) \textit{stride} (denoted as $s$): length of slide. In our dynamic data augmentation process, we employ a skewness-driven dynamic stride. The setting of \textit{stride} gets smaller for records whose labels are scarce, and gets larger for records whose labels are common, constrained by a maximum stride threshold. Formally, given a maximum stride threshold $MS$ and $m$ labels $C = \{c_1, c_2, \dots, c_m\}$, the stride of record with label $c_j$ is given by: 

\begin{equation}
\begin{centering}
\textit{s}_{c_j} = \lceil MS \times \frac{ \ | \text{records labelled} \ c_j \ |}{\max_{s=1}^m \ | \text{records labelled} \ c_s \ |} \rceil
\end{centering}
\label{eq:stride}
\end{equation}

Notice that if one ECG record has less than \textit{windows\_size} length, we pad it with zeros at the end. Therefore, a $T$-segments array $\bm{X}^{(i)} = [\bm{x}_1^{(i)}, \bm{x}_2^{(i)}, \dots, \bm{x}_T^{(i)}]$ can be generated from an ECG record $\bm{x}^{(i)} \in X$ ,  where $t \in \{1,2,\dots,T\}$-th segment $\bm{x}_t^{(i)}$ is a continuous real-valued vector $[x_{t1}^{(i)}, x_{t2}^{(i)}, \dots, x_{tw}^{(i)}] \in \mathbb{R}^w$ and $w$ is the window size. The label $y^{(i)}$ of $i$-th record $\bm{x}^{(i)}$ is assigned to all the segments in its augmented segments array $\bm{X}^{(i)}$. For clarity, we denote a segments labels set as $Y^{(i)} = [y_1^{(i)}, y_2^{(i)}, \dots, y_T^{(i)}]$ for each ECG sequence $\bm{x}^{(i)}$ where $y_t^{(i)}$ is the corresponding label for $t$-th segment $\bm{x}_t^{(i)}$.

\subsection{Multi-view Residual-Convolution-Recurrent Neural Network}\label{sec:rcr-net}
\begin{figure}[tbp]
\begin{centering}
\includegraphics[width=0.42\textwidth]{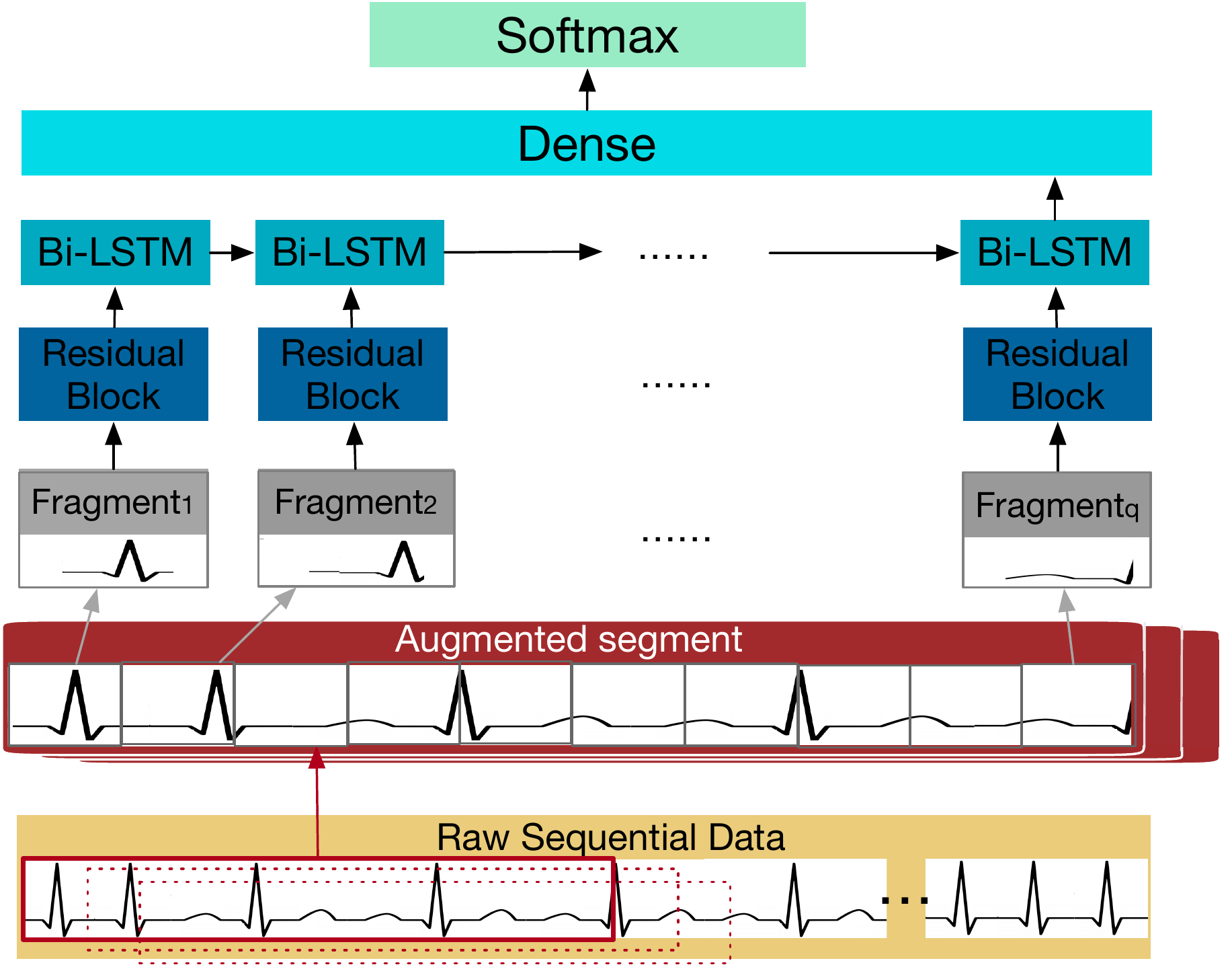}
\caption{The high-level architecture of RCR-net model}
\label{fig:RCR-net}
\end{centering}
\end{figure}

Recently, DNN methods have achieved great success in AF detection \cite{wu2016novel,majumdar2017robust,kiranyaz2015real,hannun2019cardiologist}. To build a more general deep learning method which could automatically capture anti-noise characteristics for a wider range of arrhythmias, we want to extract and integrate both local heartbeat-level characters and long-term rhythm-level trend from an ECG record, namely multi-view deep features. From this point of view, we design a multi-view residual-convolution-recurrent neural network architecture (RCR-net) to benefit from both the power of automatic feature extraction using CNN and the long-term trend captured using RNN. The high-level architecture of the RCR-net is shown in Figure \ref{fig:RCR-net}. The network takes augmented ECG segments as input, and outputs the predictions. 

In detail, RCR-net consists of a 33-layer stacked residual block \cite{he2016deep}, 1-layer recurrent block and 1-layer fully connected block. The residual block layers aim to automatically extract more effective local heartbeat-level features by constructing a very deep model via residual connections between blocks. The recurrent layer is designed to capture underlying rhythm-level structure in ECG data. Here, we use Bi-directional Long-Short Term Memory (Bi-LSTM) cells which can capture long term trend by utilizing a gated architecture. Then, the predictions are made by a fully connected layer and a softmax layer. Finally, we compute cross-entropy loss for objective function, and optimize the loss for training the neural network.

\subsection{K-margin-based Diagnosis Model} \label{sec:build}
\begin{figure}[tbp]
 \centering
 \includegraphics[width=8cm]{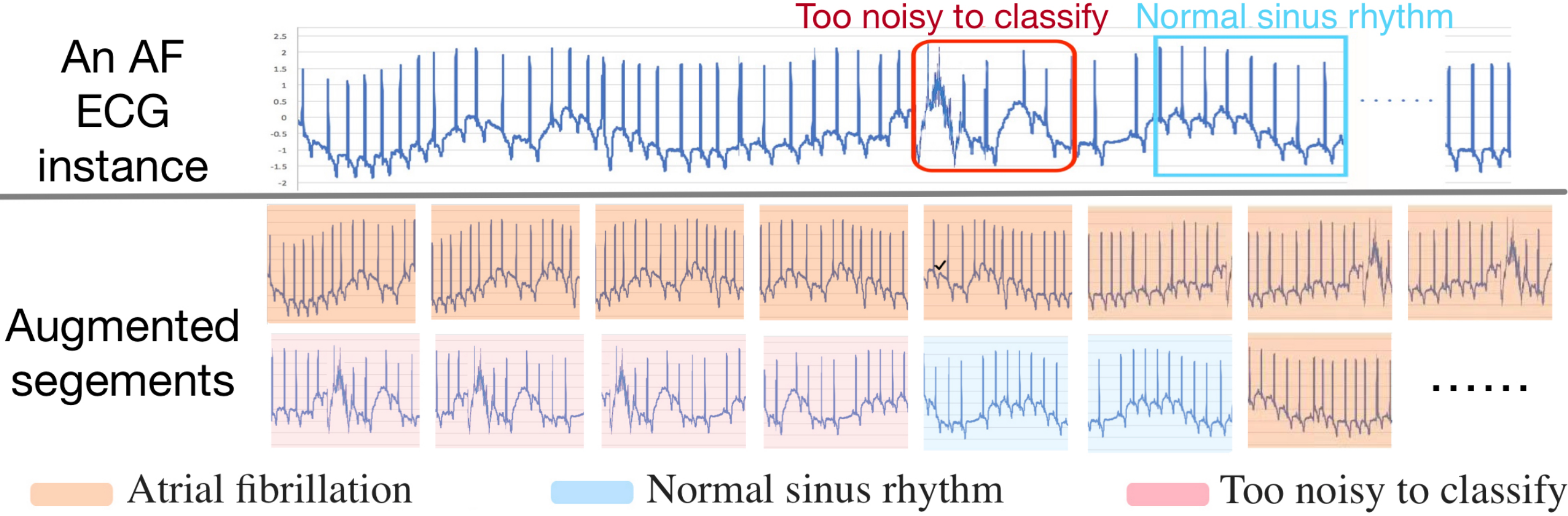}
 \caption{An example of ``hard'' segments with noisy labels caused by data augmentation process.}
 \label{fig:af_niosy}
\end{figure}

Skewness-driven dynamic data augmentation is essential to boost the model performance by handling the problems of data inadequacy, class imbalance and high computing complexity in ECG analysis. However, as shown in Figure \ref{fig:af_niosy}, it inevitably generates ``hard'' ECG segments due to noisy labels. Therefore, to enhance the robustness of RCR-net, we propose to compute cross-entropy objective function of only a portion of the selected ECG segments.

In order to select the proper segments for computing cross-entropy, we first propose a multi-class uncertainty variant measurement called least uncertainty margin. Given a trained RCR-net model, one can get all the predictions of augmented ECG segments for each ECG record. Then, we define the uncertainty margin of $t$-th segment $\bm{x}_t^{(i)}$ for the given ECG record $\bm{x}^{(i)}$ under trained RCR-net model as: 

\begin{equation}
\begin{centering}
Margin(\bm{x}_t^{(i)}) = P(\ddot{\hat{y}}_t^{(i)} | \bm{x}_t^{(i)}) \ - \ P(\dot{\hat{y}}^{(i)}_t | \bm{x}_t^{(i)})
\end{centering}
\end{equation}

where $\dot{\hat{y}}_{t}^{(i)}$ and $\ddot{\hat{y}}_{t}^{(i)}$ are the most and second-most probable predicted classes of $\bm{x}_t^{(i)}$ using the RCR-net model. Intuitively, segment with the least uncertainty margin is the most confident one for the given ECG record under the trained model, because the trained model has little doubt in differentiating between the two most probable classes. By contrast, segments with larger margins are more ambiguous. Therefore, we define the segment with least uncertainty margin of $i$-th input ECG record $\bm{x}^{(i)}$ with $T$-segments array under trained RCR-net model as:

\begin{equation}\label{eq:x_opt}
\begin{centering}
\bm{x}_{*}^{(i)}  = \mathop{\arg\min}_{\bm{x}_t^{(i)},t \in \{1,2,\dots,T\}}(Margin(\bm{x}_t^{(i)}))
\end{centering}
\end{equation}

where $\bm{x}_t^{(i)}$ is the $t$-th segment of the input ECG record $\bm{x}_i$. Least margin diagnosis aims to find the predicted classes of an ECG record by applying the most confident strategy. In a similar way, the most confident label of an input ECG record $\bm{x}^{(i)}$ under trained RCR-net model is defined as follows:

\begin{equation}\label{eq:y_opt}
\begin{centering}
y_{*}^{(i)}  = \mathop{\arg\min}_{\dot{\hat{y}}_t^{(i)},t \in \{1,2,\dots,T\}}(P(\ddot{\hat{y}}^{(i)}_t | \bm{x}_t^{(i)}) - P(\dot{\hat{y}}_t^{(i)} | \bm{x}_t^{(i)}))
\end{centering}
\end{equation}

\begin{algorithm}[tb]
    \caption{K-LABELS($\bm{X}^{(i)}, K, \phi$)}
    \label{alg:algorithm}
    \textbf{Input}: A $T$-segments array $\bm{X}^{(i)} = [ \bm{x}_{1}^{(i)}, \bm{x}_{2}^{(i)}, \dots, \bm{x}_{T}^{(i)}$] of an input ECG record $\bm{x}^{(i)}$, trained RCR-net model $\phi$\\
\textbf{Parameter}: An integer K\\
       \textbf{Output}: Top-K most confident segments $\bm{X}_{1:K}^{(i)}$  and their label predictions $\hat{Y}_{1:K}^{(i)}$
    \begin{algorithmic}[1] 
        
         \STATE $\bm{X}_{1:K}^{(i)} \gets \emptyset$, \  $\hat{Y}_{1:K}^{(i)} \gets \emptyset$, \ $k \gets K$
         \WHILE{$k > 0$}
         	\STATE $\bm{x}_*^{(i)} \gets \text{Most  \ confident \ segment \ predicting \ by} \ \phi (\bm{X}^{(i)})$ using Eq.~\ref{eq:x_opt}
         	\STATE $\hat{y}_*^{(i)} \gets \text{Most  \ confident \ label \ predicting \ by} \ \phi (\bm{X}^{(i)})$ using Eq.~\ref{eq:y_opt}
        	\STATE $\bm{X}^{(i)} \gets \bm{X}^{(i)}) \setminus \bm{x}_*^{(i)}$  \# Implementation: remove segment index of $\bm{x}_*^{(i)}$ from segment indexes list of $\bm{X}^{(i)}$
	\STATE $\bm{X}_{1:K}^{(i)} \gets \bm{X}_{1:K}^{(i)} \cup \{\bm{x}_{*}^{(i)}\}$, \  $\hat{Y}_{1:K}^{(i)} \gets \hat{Y}_{1:K}^{(i)} \cup \{y_*^{(i)}\}$
	\STATE $k \gets k - 1$
         \ENDWHILE
         \STATE \textbf{return} $\bm{X}_{1:K}^{(i)}, \  \hat{Y}_{1:K}^{(i)}$
    \end{algorithmic}
\end{algorithm}

Based on the above discussion, the proposed K-margin-based diagnosis algorithm (K-LABELS) is described in Algorithm 1. It takes an ECG record $\bm{x}^{(i)}$ as an input, and the top-K most confident segments and their labels under trained model ${\phi}$ can be identified by it. Given an ECG record $\bm{x}^{(i)}$ and its augmented segments array $\bm{X}^{(i)} = [ \bm{x}_1^{(i)},\bm{x}_2^{(i)},\dots,\bm{x}_T^{(i)}]$, it requires K iterations where K is the number of most confident segments associated with it in Algorithm 1. The top-1 most confident segment and its label needs to be found out from the segment array in iteration, and then it will be removed from the segment array. Given ${m}$ rhythm labels $\mathcal{C}$ = [${c_1}$, ${c_2}$, ..., ${c_m}$], after K iterations, the top-K most confident segments $\bm{X}_{1:K}^{(i)}$ and their label $\hat{Y}_{1:K}^{(i)}$ for the given ECG record $\bm{x}^{(i)}$ under trained model ${\phi}$ can be identified.

To avoid learning features from the noisy segments, it is expected that the model can automatically focus on the most important segments which could represent the features of the given ECG record. Intuitively, the top-K most confident segments $\bm{X}_{1:K}^{(i)}$ can be used for the training process, instead of the whole segments array $\bm{X}^{(i)}$. However, as they are selected by the RCR-net model, it is unreliable to use them for training when the model is unreliable. Therefore, for each ECG record $\bm{X}^{(i)}$, we can compute the average probabilistic prediction of all of its augmented segments:
\begin{equation}
\begin{centering}
\alpha_i \ = \frac{1}{T} \ \ (\sum_{t=1}^T P(\dot{\hat{y}}_{t}^{(i)} | \bm{x}_{t}^{(i)}) )
\end{centering}
\end{equation} 

where $T$ is the number of augmented segments within candidate $\bm{x}^{(i)}$, $P(\dot{\hat{y}}_{t}^{(i)} | \bm{x}_t^{(i)})$ is the prediction probability of $t$-th segment of $\bm{x}^{(i)}$ . If $\alpha_i > 0.5$, we select the top-K most confident segments $\bm{X}_{1:K}^{(i)}$ of given record $\bm{x}^{(i)}$ for model fine-tuning process to achieve higher consistency among the augmented segments within a record and reduce the effects of noisy segments. Otherwise, we select the top-K segments of its inverted sequence. That is because a low average probabilistic prediction of all of its augmented segments indicate a poor performance of our model. Therefore, the prediction of our model is inaccurate and unreliable, and we need more ``hard'' segment samples for training. The selected $\alpha$-segments $X^{(i)}_\alpha$ can be defined as follows:
\begin{equation}
\begin{centering}
\bm{X}^{(i)}_\alpha \ = \left\{ 
    \begin{array}{lr} 
   \bm{X}_{1:K}^{(i)}, & \text{if }  \alpha_i > 0.5\\
   \bm{X}^{(i)} \setminus \bm{X}_{1:K}^{(i)}, & \text{otherwise}
    \end{array} 
	 \right.
\end{centering}
\end{equation} 

Therefore, the task of AF detection using our K-margin-based RCR-net model can be expressed as optimizing the cross-entropy objective function:
\begin{small}
\begin{equation}
\begin{centering} 
L(X, Y) = \frac{1}{N |\bm{X}^{(i)}_\alpha|} \sum_{i=1}^N \sum_{t=1}^{|\bm{X}^{(i)}_\alpha|} \log P(Y=y_t^{(i)}| X=\bm{X}^{(i) }_{\alpha t}) 
\end{centering}
\end{equation}
\end{small}

We further transform the label predictions ${\hat{Y}_{1:K}^{(i)}}$ of the top-K most confident segments for the given ECG record $\bm{x}^{(i)}$ under trained model ${\phi}$ to a matrix $\hat{\bm{Y}}_{1:K}^{(i)} \in \mathbb{R}^{K \times m}$ as follows:
\begin{equation}
\begin{centering}
  \hat{\bm{Y}}_{k,j}^{(i)} = \left\{ 
    \begin{array}{lr} 
   1, & \text{if }  \hat{y}_{k}^{(i)} = c_j \\
    0, & \text{otherwise}\\ 
        \end{array} 
	 \right.
\end{centering}
\end{equation}

Thus, we propose a voting algorithm based on the top-K uncertainty variant measurement. To identify the most probable label of a given ECG record $\bm{x}^{(i)}$, we employ a K-margin-based diagnosis method to vote on the probability of rhythm class the ECG record ${\bm{x}^{(i)}}$ might belong to. The most probable label $\hat{y}^{(i)}$ of an ECG record $\bm{x}^{(i)}$ can be defined as follows:

\begin{equation}
\hat{y}^{(i)} = \mathop{\arg\max}_j (\sum_{1}^K \hat{\bm{Y}}_{k, j}^{(i)})
\end{equation}


Our K-margin-based RCR-net model can automatically focus on the most important segments associated with an ECG record and handle noise, as only a portion of the segments in each record participate in the learning process, by naturally exploiting expected consistency among the segments associated for each record. In our experiment, we demonstrate that our K-margin-based diagnosis can further improve the robustness and accuracy in the AF detection task. 

\section{Experiments and Results}

\subsection{Experimental Setup}

\subsubsection{Dataset}

We carry out experiments on PhysioNet/Computing in Cardiology Challenge 2017 Dataset \cite{clifford2017af}, which contains 8,528 single lead ECG records lasting from 9 s to just over 60 s sampling with 300 Hz. In this dataset, all non-AF abnormal rhythms are treated as a single class, namely Other rhythm. Thus these records are classified as 4 classes: 1) Normal sinus rhythm \textbf{N}, 2) Atrial Fibrillation \textbf{A}, 3) Other rhythm \textbf{O}, 4) Too noisy to classify \textbf{P}.

\subsubsection{Evaluation Measurements}
 Commonly used multi-class metrics such as Precision, Recall and Hamming Loss are utilized to measure the performance by evaluating how close the predicting labels are to corresponding labels given by doctors: 
\begin{itemize}
\item $Precision \ = \frac{1}{m}\sum_{c=1}^{m} \sum_{i \in \{i|y^{(i)}=c\}} \frac{\mathds{1}(y^{(i)} = \hat{y}^{(i)})}{|\{i|y^{(i)}=c\}|}$.
\item $Recall \ = \frac{1}{m}\sum_{c=1}^{m} \sum_{i \in \{i|y^{(i)}=c\}} \frac{\mathds{1}(y^{(i)} = \hat{y}^{(i)})}{|\{i|\hat{y}^{(i)}=c\}|}$.

\item $Hamming \ Loss \ = \frac{1}{N} \sum^{N}_{i = 1} \frac{\mathds{1}(y^{(i)} \neq \hat{y}^{(i)})}{m}$.
\end{itemize}

We also use the following ${F_1}$ evaluation measurements: 
\begin{itemize}
\item \textit{$F_1$ scores of each class}: Denoted as ${F_{1N}}$ for normal sinus rhythm, ${F_{1A}}$ for AF, ${F_{1O}}$ for other rhythm, ${F_{1P}}$ for noise. Detailed definitions can be found in \cite{clifford2017af}.
\item \textit{Averages of $F_1$ scores}: $F_{1NAO}=\frac{F_{1N}+F_{1A}+F_{1O}}{3}$, $F_{1NAOP}=\frac{F_{1N}+F_{1A}+F_{1O}+F_{1P}}{4}$. 
\end{itemize}

\subsubsection{Implementation Details}

In implementation, we randomly split 80\% for model training, and evaluate on remaining 20\% testing data. We evaluate the effects of various state-of-art DNN methods using the measurements given above by repeatedly running 20 times using 5-fold cross validation, and report the average results. 

We set the parameter value of window\_size  and max stride threshold ${MS}$ to be 6000 and 500 for skewness-driven dynamic data augmentation respectively. Then we set the parameter value of K and ${N\_split}$  to be 3 and 300 for training K-margin-based RCR-net using augmented data (see Section \ref{sec:augment}), and identifying the final predicting label of each ECG record by the trained model. Methods are implemented using Python 3.6.2 on TensorFlow version r1.4.

\subsection{Results}

\begin{figure}[tbp]
\begin{centering} 
\begin{minipage}[t]{8cm}%
\includegraphics[width=3.8cm]{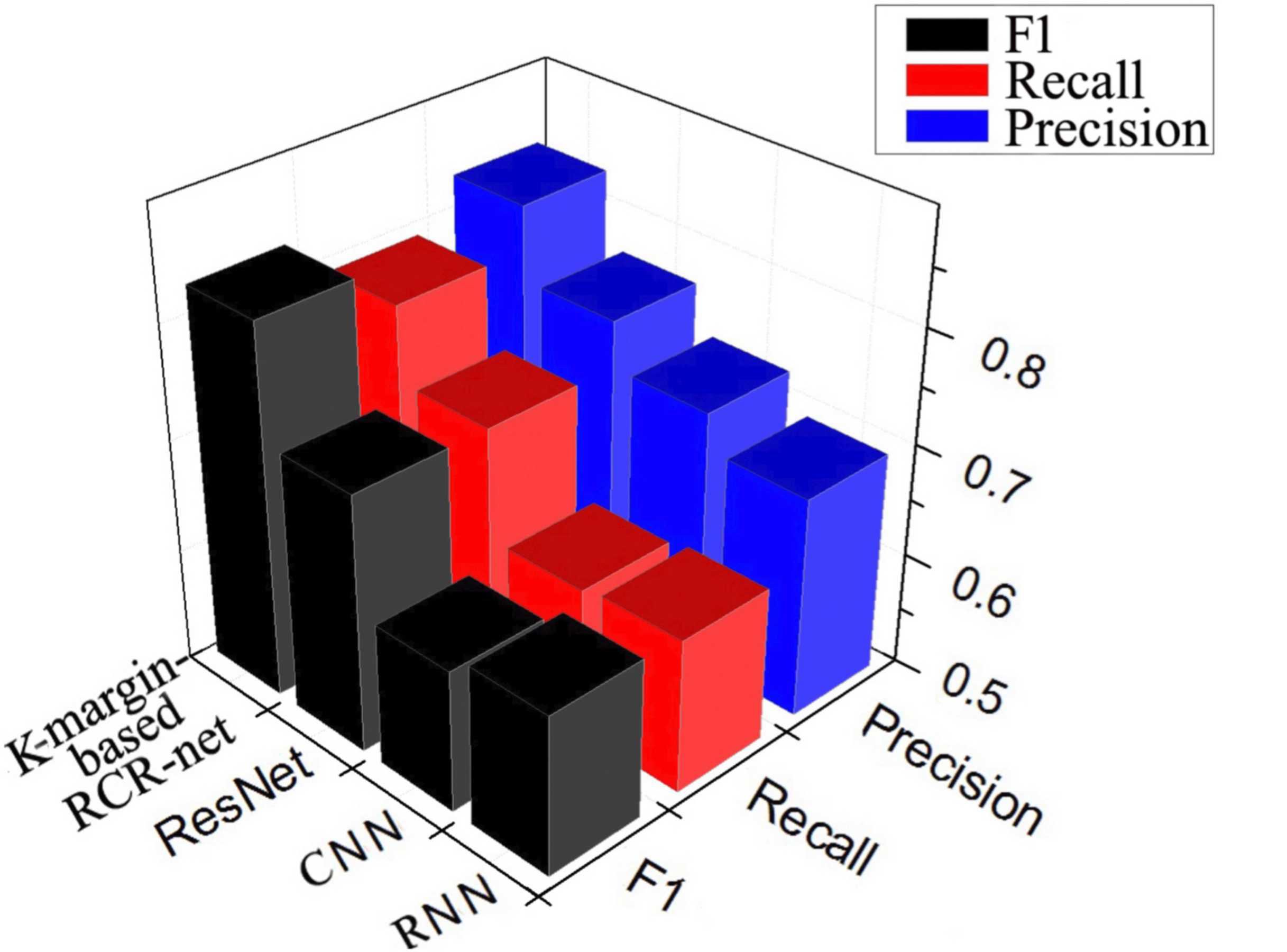}
\includegraphics[width=4.1cm]{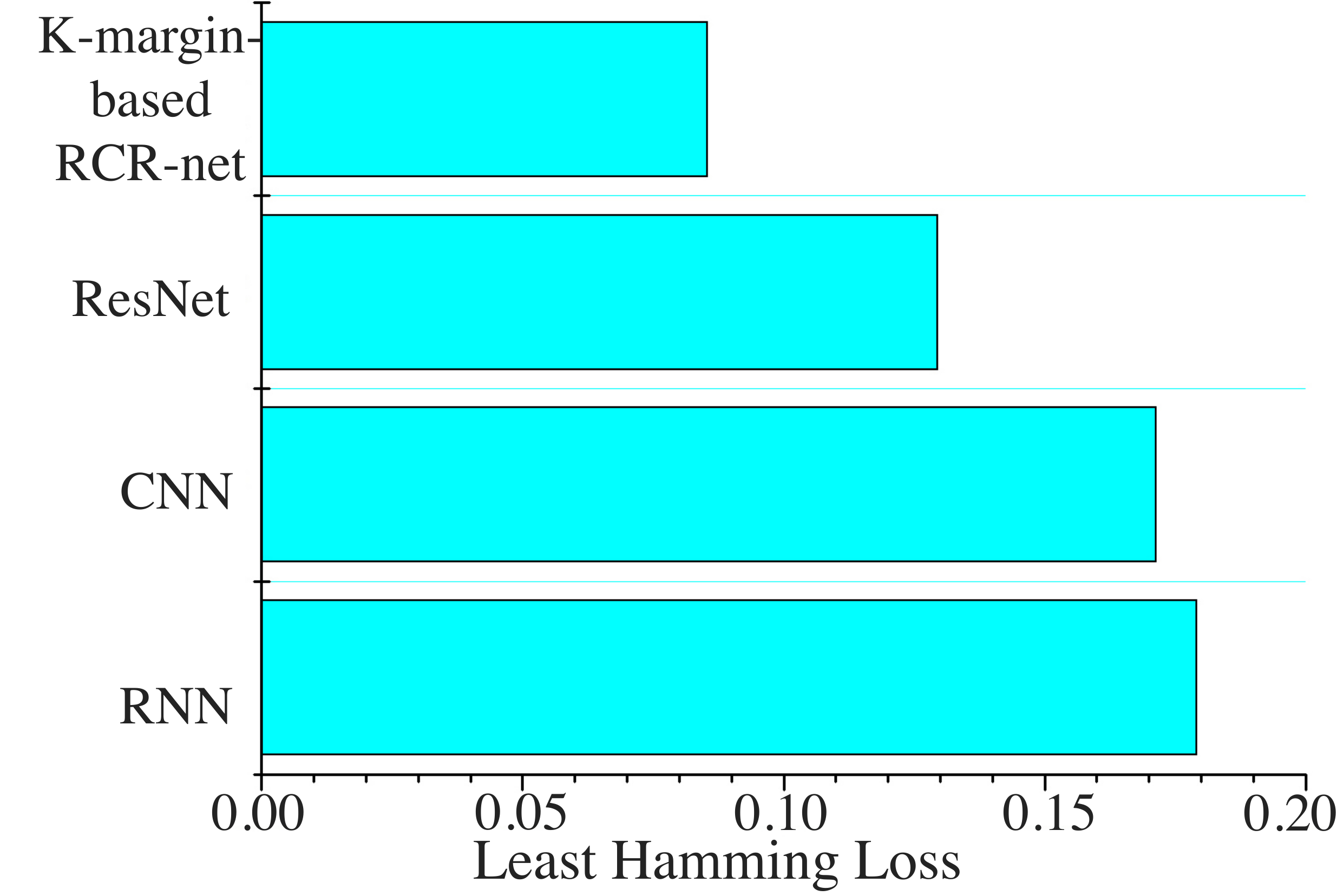}
\end{minipage}%
\caption{(Left) The result of Precision, Recall, F1 scores. (Right) The result of hamming loss scores. }
\label{fig:arf_hs}
\end{centering}
\end{figure}

\begin{figure}[tbp]
\begin{centering} 
\begin{minipage}[t]{8cm}%
\includegraphics[width=3.7cm]{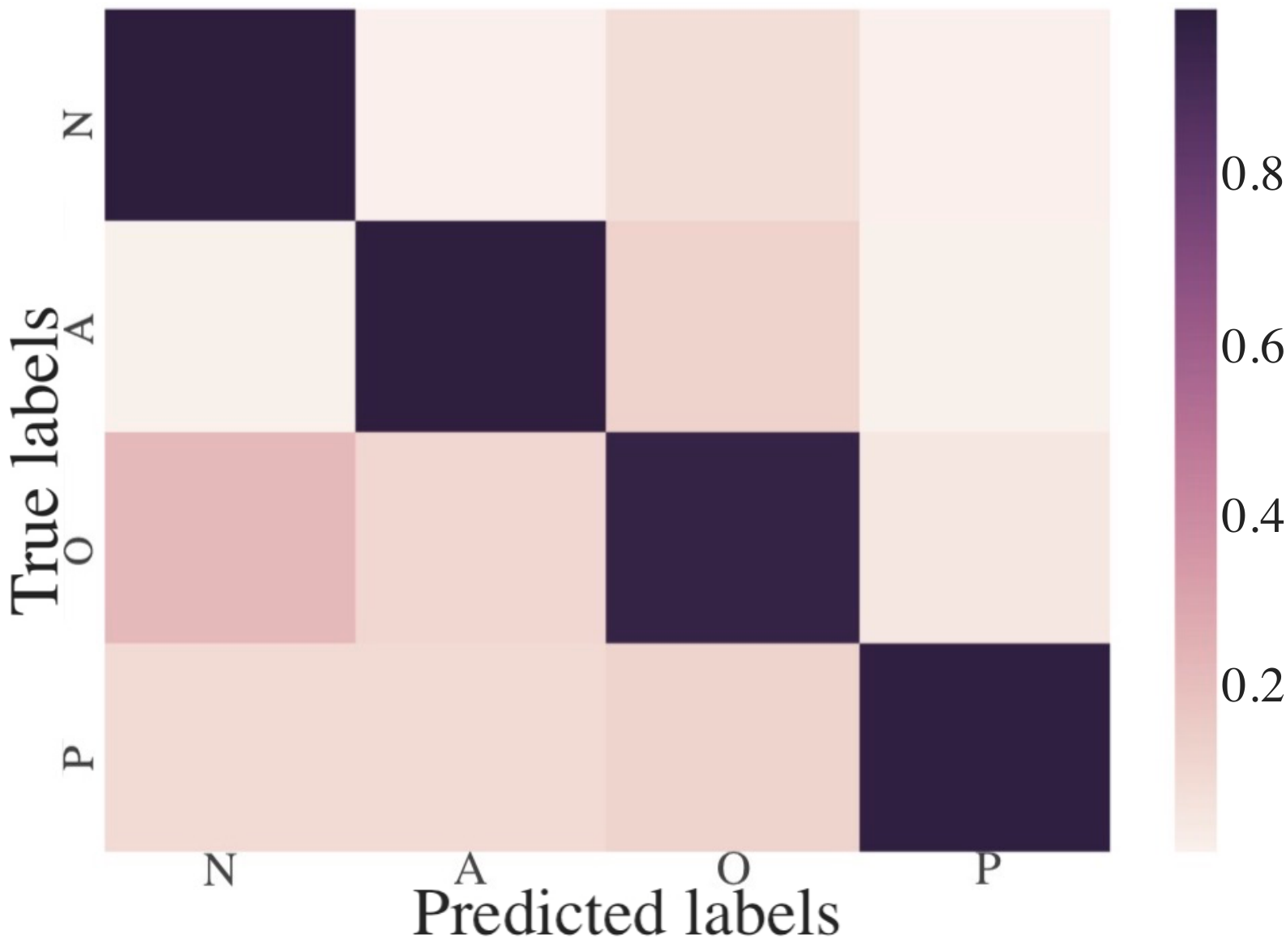}
\includegraphics[width=4.2cm]{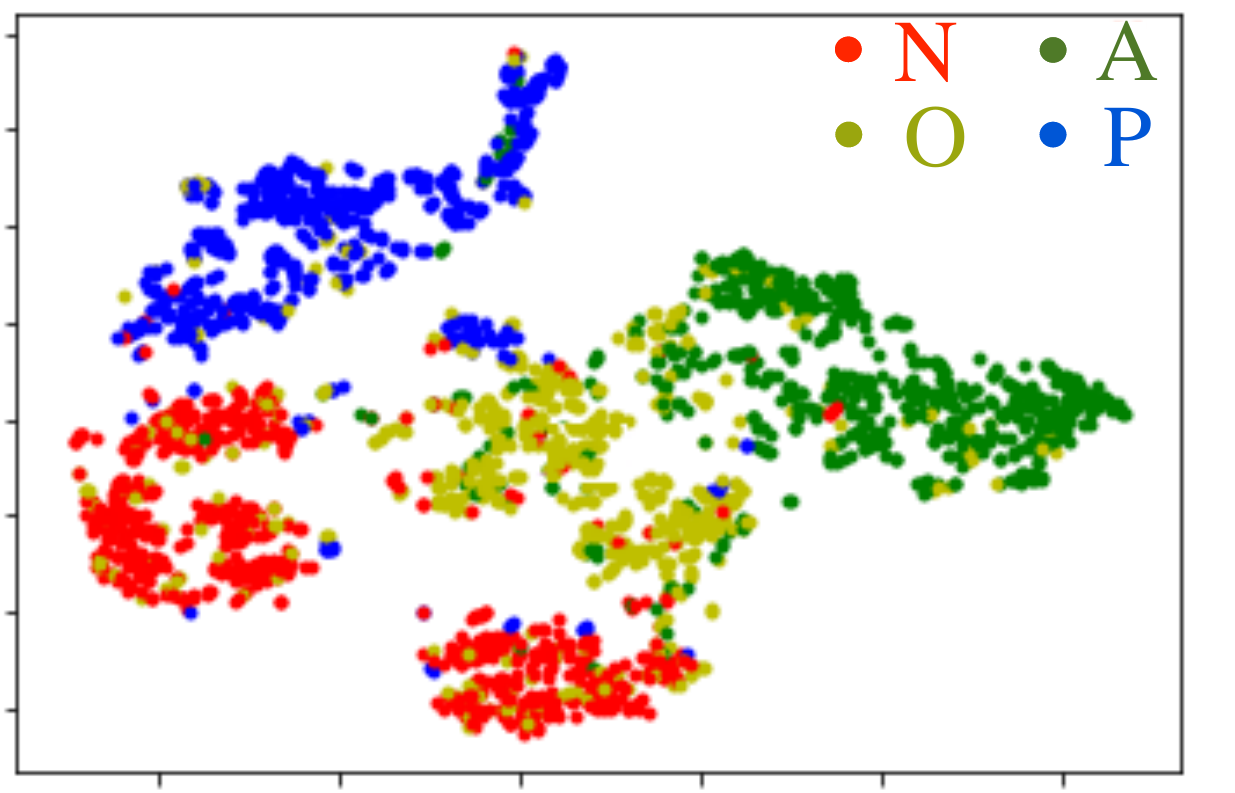}
\end{minipage}%
\caption{Confusion matrix and deep features visualization of RCR-net.}
\label{fig:cfm_df}
\end{centering}
\end{figure}

\subsubsection{Comparing with Other Methods}

First of all, we compare our method with the following deep neural network methods: 
\begin{itemize}
\item \textit{RNN} is one of the most popular DNN architecture, which allows it to exhibit temporal dynamic behavior. It has been used as main classifier for detecting AF in \cite{warrick2018ensembling}.
\item \textit{CNN} use a variation of multilayer perceptrons to automatically extract features. Recently, \cite{sodmann2018convolutional} constructed CNN architecture for AF detection without manual extracted features.
\item \textit{ResNet} allows training of very deep networks by constructing the network through modules called residual models, which can avoid the higher training error caused by naively adding the layers \cite{he2016deep}. In \cite{hannun2019cardiologist}, they have achieved accuracies comparable to or higher than cardiologists for classifying 14 different types of arrhythmia by employing the ResNet as the main classifier.
\end{itemize}

Results are shown in Figure \ref{fig:arf_hs}. The left part shows precision, recall and ${F_{1NAOP}}$ results. We can see that the proposed K-margin-based RCR-net method with 0.8125 $F_{1NAOP}$ score outperforms all state-of-the-art deep learning methods for AF detection task by $6.8\%$. The method of AlexNet and VGG are not effective, because it is difficult to construct very deep networks which will cause information loss and feature loss. The ResNet method works poorly because feature mapping simply aggregates all the automatic extracted local characters without considering any context information of overall long-term trend, which is constrained by the expressiveness of the model.

From the view of generalization ability of predictor (Hamming Loss), our K-margin-based RCR-net also performs better than others in terms of 38\% - 50\% fraction of labels that are incorrectly predicted, as shown in Figure \ref{fig:arf_hs} (right).

\begin{table*}[tbp]
\centering
\resizebox{0.75\textwidth}{!}{
\begin{tabular}{l|llll|ll}
\toprule  Method & \textbf{${F_{1N}}$}	& \textbf{${F_{1A}}$}	& \textbf{${F_{1O}}$}	& \textbf{${F_{1P}}$}	& \textbf{${F_{1NAO}}$}	& \textbf{${F_{1NAOP}}$} \\ \midrule
Ours	& 0.9061	& \textbf{0.8712}	& 0.7166	& \textbf{0.7561}	& \textbf{0.8313}	& \textbf{0.8125}  \\ 
\cite{hong2017encase}	& \textbf{0.9117}	& 0.8128	& \textbf{0.7505}	& 0.5671	& 0.8250	& 0.7605  \\ 
\cite{zihlmann2017convolutional}	& 0.9090	& 0.8221	& 0.7319	& 0.5676	& 0.8210	& 0.7577  \\ 
\cite{xiong2017robust}	& 0.9031	& 0.8203	& 0.7310	& 0.5251	& 0.8181	& 0.7449  \\ 
\cite{schwab2017beat}	& 0.9062	& 0.7385	& 0.7165	& 0.4751	& 0.7871	& 0.7091  \\ 
\cite{andreotti2017comparing}	& 0.8923	& 0.7553	& 0.6715	& 0.4843	&0.7730	& 0.7009 \\ 
\cite{jimenez2017atrial}	& 0.8990	& 0.7708	& 0.6944	& 0.4121	& 0.7881	& 0.6941  \\ 
\cite{zihlmann2017convolutional}	& 0.8884	& 0.7647	& 0.6686	& 0.4220	& 0.7739	& 0.6859  \\ 
\cite{stepien2017classification}	& 0.8973	&0.7012	& 0.6420	& 0.4377	&0.7468	& 0.6696  \\ 
\cite{chandra2017atrial}	& 0.8600	& 0.7300	& 0.5600	& N.A.	& 0.7167	& N.A.  \\ \bottomrule
\end{tabular}
}
\caption{Comparing our method with other deep learning methods in the PhysioNet/Computing in Cardiology Challenge 2017.}
\label{exp:f1InChallenge2017}
\end{table*}

\begin{figure*}[tbp]
\begin{centering} 
\resizebox{0.9\textwidth}{!}{
\begin{minipage}[t]{18.8cm}
\includegraphics[width=7.3cm]{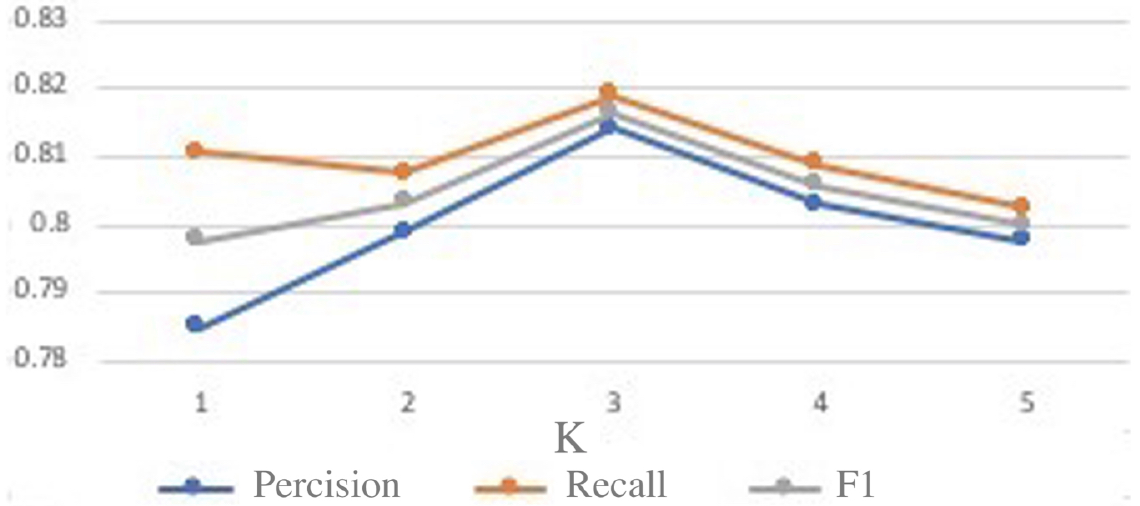}
\includegraphics[width=5.4cm]{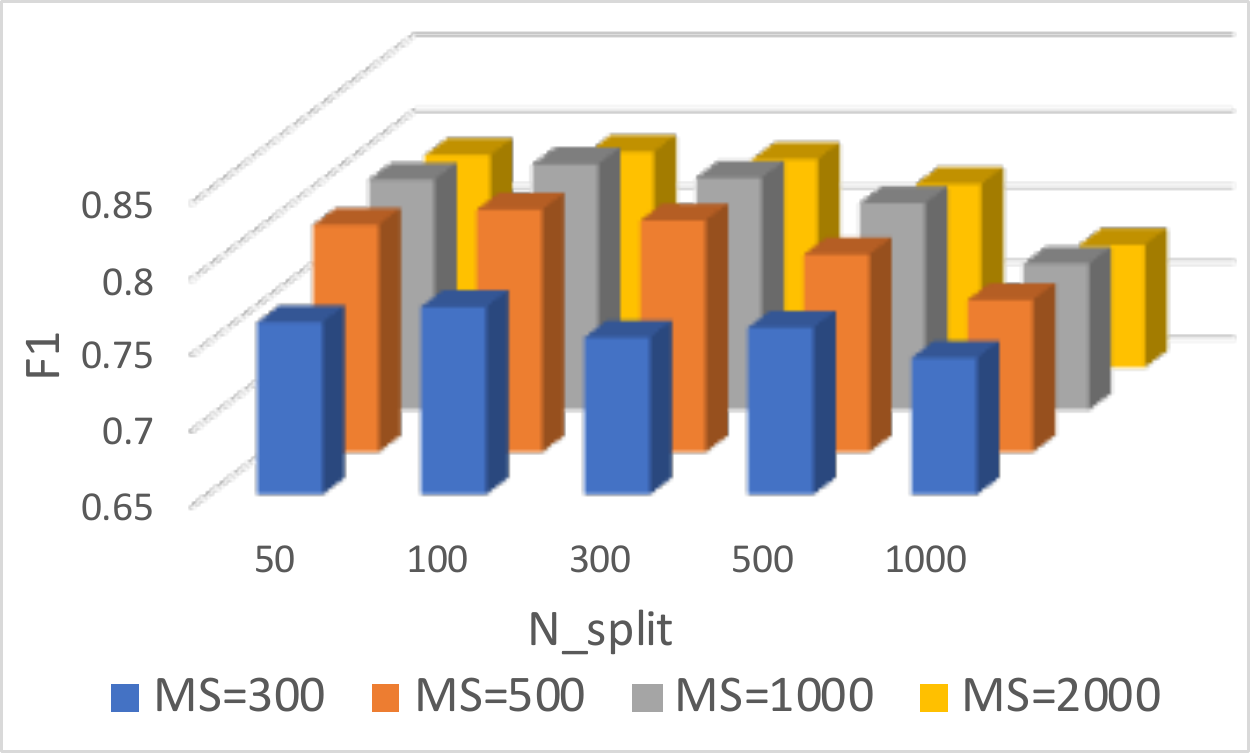}
\includegraphics[width=5.4cm]{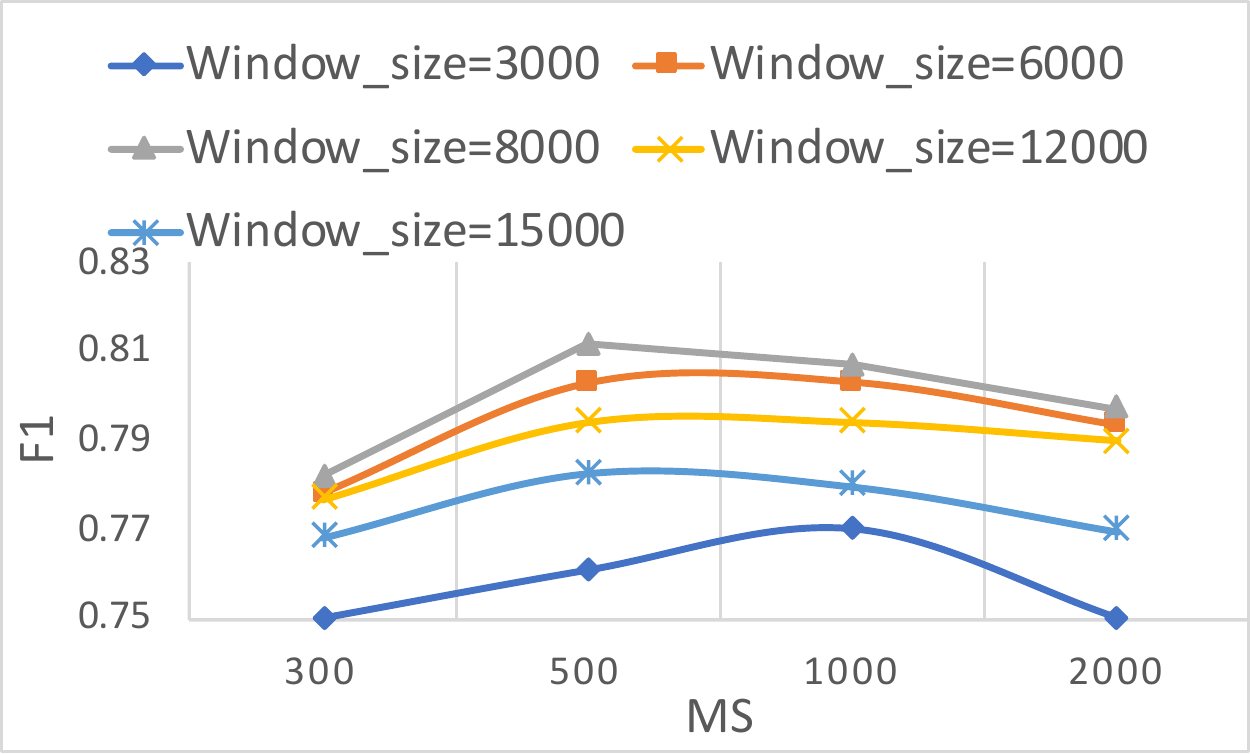}
\end{minipage}
}
\caption{${F_1}$ score results of K-margin-based RCR-net for different parameters.}
\label{fig:parameters}
\end{centering}
\end{figure*}

Besides, Table \ref{exp:f1InChallenge2017} also demonstrates our $F_1$ score and nine DNN methods employed in the Challenge 2017. We can see that automatically focusing on the most important segments (with a $6.8\% - 21.3\%$ higher ${F_{1NAOP}}$ score when comparing it with other DNN methods) is an effective way to improve prediction performance (especially on handling noisy segments and identifying Too Noisy To Classify records).

Moreover, Figure \ref{fig:cfm_df} shows detailed performance of our method. The left part of Figure \ref{fig:cfm_df} shows a confusion matrix of the model predictions on the test set. Often mistakes made by the model are understandable. For example, many records of Other Rhythm are confused with Normal Sinus Rhythm which makes sense given that it is very difficult to distinguish between the normal sinus rhythm and some arrhythmias in the ECG records. Confusing Other Rhythm and Too Noisy To Classify also makes sense, as this can be subtle to detect especially when their ECG morphologies are similar or when noise is present. The right part of Figure \ref{fig:cfm_df} shows deep features learnt using the K-margin-based model on the test set. This shows whether the learned deep features effectively capture the latent relationships among the given four classes. Obviously, we can come to the conclusion that the learnt deep features are highly distinguished and representative.

\subsubsection{Hyper-parameters Analysis}

Now we further evaluating the performance of our method in different hyper-parameter settings. 

The effects of threshold K are shown in the left part of Figure \ref{fig:parameters}. We display the accuracy results of our K-margin-based RCR-net under different threshold K in the range of $[1, 3]$ and how the accuracy (Precision, Recall and ${F_1}$ score) raises dramatically. However, it drops when the K-margin threshold K is larger than 3. The reason is that the accuracy of our K-margin-based RCR-net model depends on both the number of segments engaged in the algorithm and the confidence of their labels. As K determines how many segments of an ECG record are engaged in the algorithm. Therefore, a little bit larger value of threshold K may result in higher accuracy. While the confidence of labelling accuracy decreases when K gets larger, as we select top-K segments from the most confident one to a less confident one. It means that higher K may also result in smaller confidence of labels engaged in the learning and voting phase, and therefore it would decrease the performance of our K-margin-based RCR-net model.

In addition, we evaluate the effects of three main parameters (windows\_size, maximum stride threshold $MS$ and N\_split). In the middle part of Figure \ref{fig:parameters}, we display the accuracy results of our K-margin-based RCR-net model under different parameter ${N\_split}$  and MS in the range of $[50, 1000]$ and $[300, 2000]$ respectively. We can see that the model performs better by splitting ECG segments into smaller fragments other than larger fragments, as it could provide more fragments for recurrent neural network layer to capture rhythm-level trend features. However, the length of a fragment decreases with the increasing of the number of fragments which would cause the loss of some critical local heartbeat-level features.

Furthermore, as shown in the right part of Figure \ref{fig:parameters},  when the model parameters of \textit{window\_size} and MS vary within the range of $[3000, 8000]$ and $[300, 2000]$ respectively, as smaller parameters increase the augmented training data, the K-margin-based RCR-net performs better. However, the performance decreases when using very small ones, as it can only provide a very narrow view of local beat-level features.

\section{Conclusion}

In this paper, we propose a K-margin-based RCR-net for AF detection. The experimental results demonstrate that the proposed method with 0.8125 $F_{1NAOP}$ score outperforms all state-of-the-art deep learning methods for AF detection task by $6.8\%$. A possibly rewarding avenue of future research is to consider multi-modality data input and more fine-grained output categories to improve the model performance and apply it in a more practical situation.

\bibliographystyle{named}
\bibliography{ijcai19}

\end{document}